\documentclass[a4paper,11pt]{article}
\usepackage{pos}
\usepackage{lineno}

\title{Physics Perspectives with the ePIC Far-Forward and Far-Backward detectors}
 \ShortTitle{Forward Physics with the ePIC experiment}

\author*[a,b]{Michael Pitt}

\affiliation[a]{Ben-Gurion University of the Negev, Department of Physics,
  Beer-Sheva, Israel}

\affiliation[b]{The University of Kansas, Department of Physics,
Lawrence, USA}

\emailAdd{michael.pitt@cern.ch}

\abstract{The forthcoming Electron--Ion Collider (EIC), which is expected to commence operations in the early 2030s, has already reached several significant milestones on its path toward completion. The core of the EIC physics program is the 3D imaging of partonic structures in protons and nuclei. The experimental detector setup required to enable this primary objective utilizes ``far-forward'' (FF) and ``far-backward'' (FB) detectors positioned downstream in the hadron-going and electron-going directions, respectively, from the interaction point of the EIC. The primary purpose of the FB detectors is to monitor luminosity and measure scattered electrons in collisions in the EIC, while the array of FF detectors is used to tag and reconstruct both charged and neutral particles that scatter at small angles. These detectors also enable a broader physics program than was initially envisioned, enhancing the EIC's research potential. The expanded capabilities have been a prime focus for engaging the broader nuclear physics community to build a robust groundwork for the EIC. In these proceedings, we will describe the FF/FB detectors and review the advanced forward physics program facilitated by them at the EIC.}

\FullConference{31st International Workshop on Deep Inelastic Scattering (DIS2024)\\
 8–12 April 2024\\
Grenoble, France\\}

\begin{document}
\maketitle

\section{Introduction}
The ePIC experiment, which is scheduled to begin in the early 2030s at the future Electron--Ion Collider (EIC) at Brookhaven National Laboratory (BNL), is poised to deepen our understanding of the fundamental structure of visible matter. The primary objectives of the scientific mission of the EIC, as outlined in a 2018 report by the National Academy of Sciences \cite{NAS:report}, include the determination of the full 3D momentum and spatial structure of nucleons, with a focus on understanding the gluon density and saturation phenomena. Additionally, the ePIC experiment aims to elucidate how the mass and the spin of nucleons and other hadrons arise from strong interactions.

The ePIC experiment will comprise a central 10-meter-long cylindrical barrel detector, covering a rapidity range from -4 to 4. An additional array of detectors that extends approximately 50 meters in the forward ($\eta>4$) and backward ($\eta<-4$) directions will be incorporated and is essential for achieving key objectives, such as measuring luminosity, tagging low-Q$^2$ scattered electrons, and measuring both prompt and secondary particles in the rapidity range of $\eta>4$. The extended detector array also expands the scope of the physics program beyond what was initially envisioned, significantly enhancing the EIC's research potential. The following section will delve into the details of the forward and backward detector arrays and explore some of the new physics opportunities made possible by these detectors.

\section{The extended ePIC detector}

\subsection{The Far-Backward detectors}
\subsubsection*{Luminosity monitor}
Precise cross-sectional measurements place stringent requirements for luminosity determination at the EIC. The luminosity monitor is set to measure the luminosity from the electron--ion elastic bremsstrahlung process $ep\to e\gamma p$ with a precision better than 1\%, as this process has a very large cross-section~($\sim$mb)~\cite{Haas:2010bq}. The detector concept follows a detector design similar to that used at ZEUS, HERA \cite{Helbich:2005qf}, and it employs two distinct methods for counting bremsstrahlung photons: photon conversion into $e^+e^-$ pairs for precise DIS cross-sectional measurements and direct (non-converted) photon detection for monitoring instantaneous collider performance.

Bremsstrahlung photons deviate from the electron beam and exit the beampipe through a 1-cm-thick aluminum window, which acts as an energy filter for them. However, this necessary thickness of the exit window inevitably causes some photon pair conversions, which are then eliminated by a sweeper magnet positioned after the exit window. One percent of the bremsstrahlung photons are converted into $e^+e^-$ pairs using a 1-mm-thin aluminum foil and are then directed into a pair spectrometer by an adjustable spectrometer magnet. The pair spectrometer comprises a tracking layer (AC-LGAD \cite{Mandurrino:2020ukm}) with a spatial resolution of 20~$\mu m$, followed by a scintillating fiber calorimeter with a thickness of 23$X_0$. The remaining bremsstrahlung photons are measured in a direct-photon calorimeter. 
 
\subsubsection*{Low-Q2 taggers}
\begin{figure}
    \centering
    \includegraphics[width=2.5in]{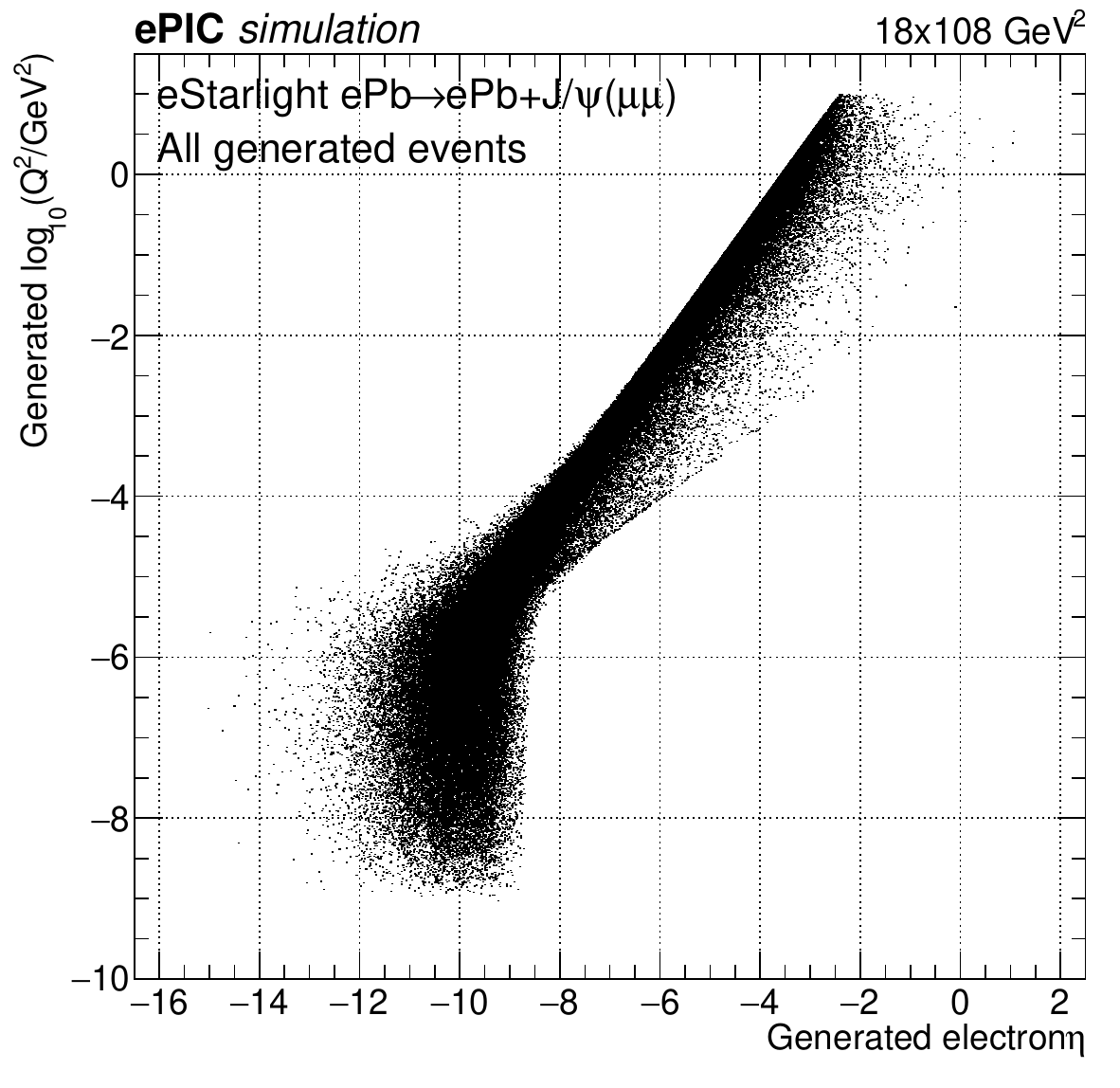}
    \includegraphics[width=2.5in]{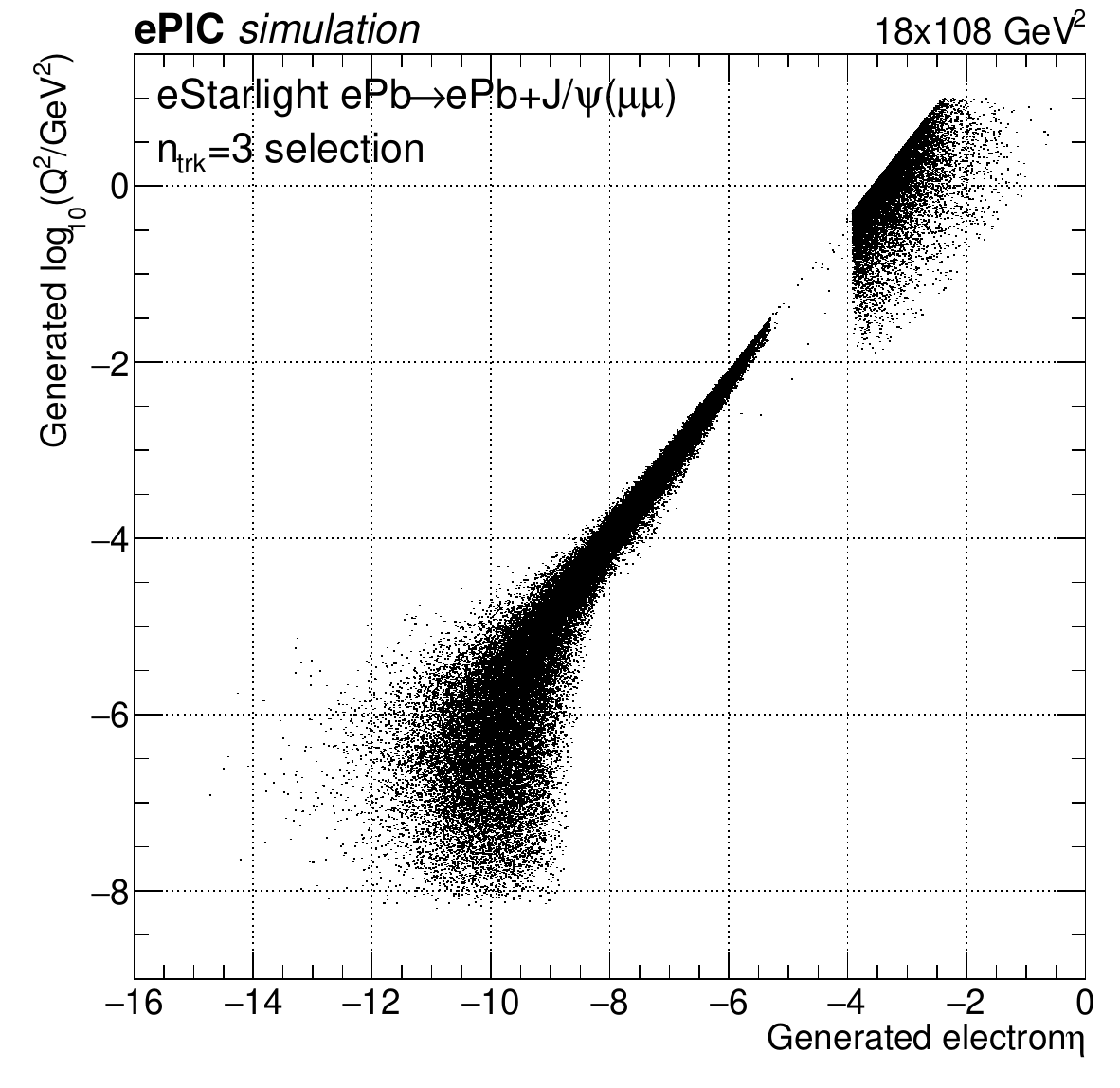}
    \caption{The electron kinematics as a function of rapidity and $Q^2$ in coherent $J/\psi$ events, simulated using the eStarlight event generator~\cite{Lomnitz:2018juf} for $ePb$ collisions at a center-of-mass energy of $18~\textrm{GeV}\times 108~\textrm{GeV}$ and $Q^2<10$~GeV$^{2}$. The results are shown for all generated events (left) and for  events where exactly three tracks are reconstructed---either all three tracks in the central detector or two tracks in the central detector with one track in the low-$Q^2$ tagger (right).}
    \label{fig:lowQ2}
\end{figure}

Photon virtuality ($Q^2$) is closely related to the scattering angle of the outgoing electron in $eA$ collisions. In these interactions, the central detector, which covers a rapidity range down to $\eta=-4$, has a high acceptance for outgoing electrons with $Q^2>1~\text{GeV}^2$. However, scattered at smaller angles, electrons with low $Q^2$ values tend to escape detection. Enhancing the detector's ability to cover these small angles would not only allow to probe a wider range of kinematic regions but would also provide valuable insights into processes involving quasi-real photons in the $Q^2$ range between 10$^{-3}$ to 10$^{-1}$. Figure \ref{fig:lowQ2} shows the correlation between the rapidity of scattered electrons and $Q^2$ at the generator level before and after applying track reconstruction using the ePIC detector simulation. While there is high acceptance for $Q^2$ values below 10$^{-3}$, this very low region will likely be dominated by background tracks, which will be the focus of future studies. 

\subsection{The Far-Forward detectors}
\begin{figure}
    \centering
    \includegraphics[width=5in]{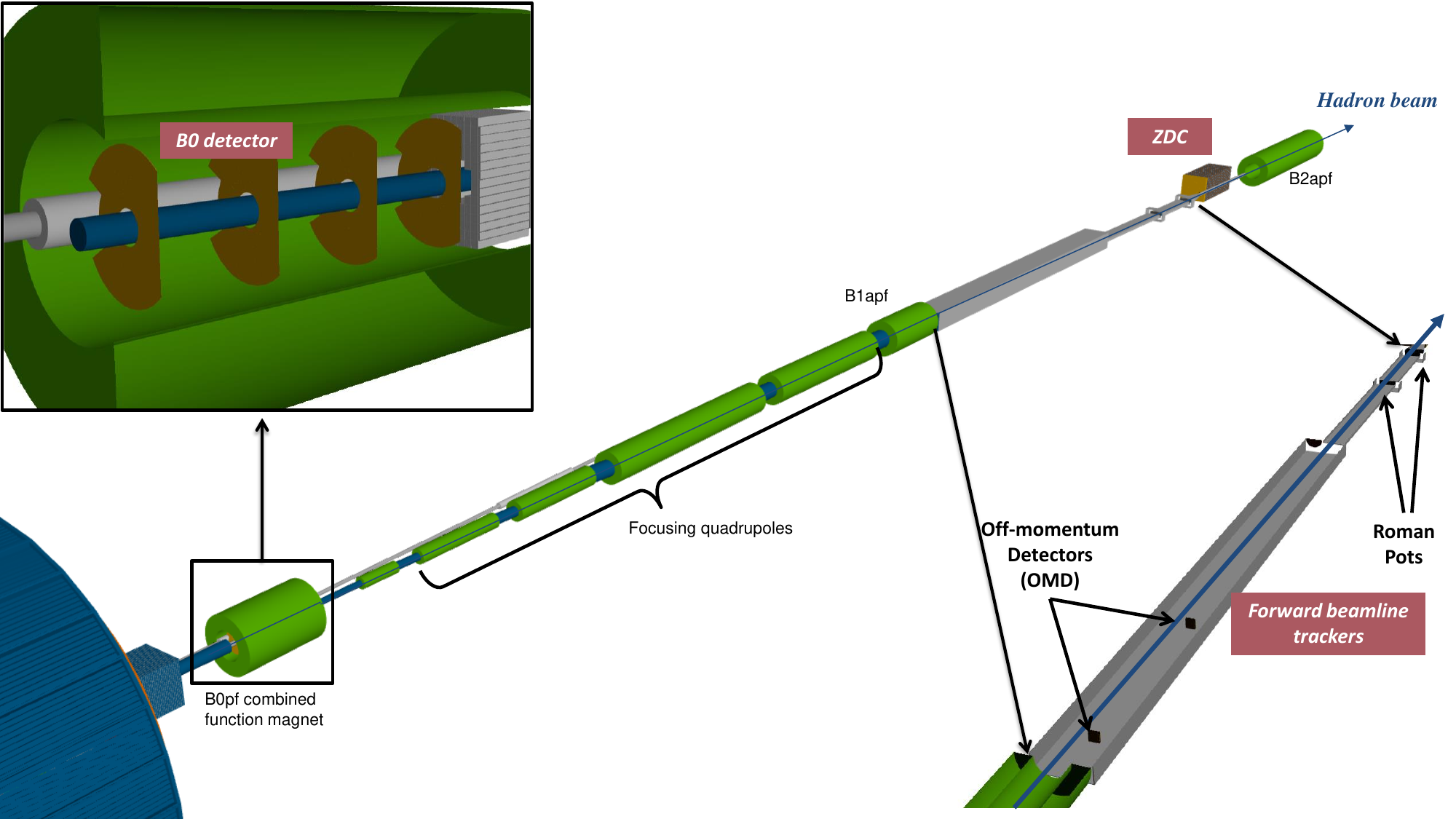}
    \caption{The layout of the ePIC far-forward region features the far-forward detector array, which consists of three subsystems: the B0 detector, the forward beamline trackers, and the zero-degree calorimeter.}
    \label{fig:FFarray}
\end{figure}

All processes of interest at the EIC are associated with the production of very forward particles, necessitating strong detection capabilities for hadrons and photons in the far-forward region ($\eta>4$). The far-forward (FF) array includes various detector concepts tailored to meet the demands of the physics program, such as calorimetry for neutrons and photons, silicon sensors for charged particle tracking and timing, and specialized detectors such as Roman pots for detecting protons or nuclear fragments that are very close to the beam. These elements are discussed in detail in this subsection. The layout of the FF detectors is shown in Figure \ref{fig:FFarray}.
    
\subsubsection*{The B0 detector}
The B0 detector consists of four evenly spaced silicon tracker layers based on AL-LGAD technology, along with an electromagnetic calorimeter (EMCAL) made up of $2\times 2\times 20$~cm$^3$ PbOW$_4$ crystals, all positioned within the bore of the B0 dipole magnet. The entire B0 detector subsystem is designed to measure particles produced at scattering angles between 5.5 and 20 mrad. The current design features a very low material budget in the rapidity range of $5 < \eta < 5.5$, allowing for high acceptance of photons across a broad energy spectrum (greater than 50 MeV), including low-energy de-excitation photons. The EMCAL is planned to achieve an energy resolution of 6--7\% and a position resolution of approximately  3~mm. With more than one interaction length, the EMCAL has a detection efficiency of over 50\% for neutrons, making it suitable for veto studies. The B0 tracker can measure forward protons with a momentum resolution $dp/p$ of 2--4\%.

\subsubsection*{Forward beamline trackers}
The very forward tracker is composed of two detectors:  off-momentum detectors (OMDs) and  Roman pots (RPs). Charged particles with lower magnetic rigidity experience greater deflection in the magnetic fields compared with nominal beam particles, leading to a displacement from the beam center. In $eA$ collisions with $A>1$, most of the accelerated spices have $A/Z\sim 2$ when protons are emitted from the ion breakup, and they will carry half of the energy required to keep the protons in orbit, resulting in a large displacement from the beam center. The OMDs, which are positioned just after the B1 dipole magnet, measure protons and other charged particles with a beam rigidity ($x_L$) ranging from 30\% to 60\%. Protons and other charged particles with a beam rigidity above 60\% are detectable in the RP detectors, which are placed just a few millimeters from the hadron beam in both the vertical and horizontal directions. The detectors also provide acceptance for scattered protons for up to 5~mrad in all directions. 

Both the OMDs and RPs consist of two AC-LGAD-based tracking planes that are spaced two meters apart and are capable of measuring both the hit position and the local scattering angle between two planes. The position of a charged particle and its scattering angle  at a given distance from the interaction point (IP) can be determined by considering its kinematics at the IP and applying a transformation matrix, which is evaluated for different beam optics configurations and particle rigidities at different distances from the IP. By inverting this equation, the track coordinates at the detector plane can be translated into the particle's kinematics at the interaction point.

\subsubsection*{Zero degree calorimeter}
Neutral particles produced at scattering angles below 5~mrad propagate in a straight line through an exit window to a dedicated zero-degree calorimeter (ZDC), which is positioned before the B2 dipole magnet, as illustrated in Figure \ref{fig:FFarray}. The ZDC design includes both electromagnetic and hadronic calorimeters. The electromagnetic section consists of 20-cm-long LYSO or PbOW$_4$ crystals, which are optimized for detecting soft photons, while the hadronic section will be similar to the ePIC forward hadron calorimeter \cite{Klest:2024xlm}. The reconstruction of neutral particles utilizes a machine-learning-based approach using the HEXPLIT algorithm \cite{Paul:2023okc}, which meets the detector requirements of an energy resolution of $\Delta E=50\%/\sqrt{E}\oplus 5\%$  and a position resolution  of $\Delta\theta=3\textrm{mrad}/\sqrt{E}\oplus 5\%$ for neutrons, as well as an energy resolution of $\Delta E=5\%/\sqrt{E}\oplus 3\%$  and a position resolution of 0.5--1~mm  for photons, where $E$ represents the energy deposition in units of GeV.

\section{Physics perspectives}

The far-forward and far-backward detectors at the EIC expand the scope of the physics program beyond the initial expectations, enhancing the research potential of the ePIC experiment. These detectors play a crucial role in several key processes. The deeply virtual Compton scattering process and deeply virtual meson production are essential for imaging the transverse spatial distribution of quarks and gluons within a proton during $ep$ collisions. These processes rely on the precise detection of the intact proton, which is often detected in the far-forward region. In $ed$ collisions, short-range correlations between nucleons can be investigated by tagging proton and neutron~\cite{Tu:2020ymk}. This method provides insights into the dynamics arising from gluons in the low-x region and how these dynamics depend on the internal configuration of nucleons. The Sullivan process offers a unique opportunity to study the form factors and structure functions of pions and kaons. This process involves the production of an outgoing baryon at very forward rapidities. Saturation effects and the internal structure of nuclei can be explored through diffractive production processes or coherent vector meson production. These phenomena are particularly sensitive to the distribution of gluons within nuclei and provide valuable information about the onset of gluon saturation. The structure of free neutrons and the EMC effect can be investigated by tagging spectators in the interactions of light nuclei, such as  $e+3\textrm{He}$ collisions. Spectator tagging allows for the isolation of specific interaction channels, providing a clearer view of neutron structure and the modification of nucleon structure within nuclei. In interactions of heavy nuclei, the high acceptance for soft photons in the far-forward detectors enables groundbreaking measurements of ion de-excitations during coherent $eA$ scattering. These measurements provide insights into the de-excitation processes and the energy dissipation in excited nuclear states that emerge from high-energy collisions.

This section elaborates on two of these processes in more detail, illustrating the enhanced capabilities provided by the far-forward detectors at the EIC.

\subsubsection*{Coherent Vector Meson Production}
One of the golden measurements at the EIC is the study of coherent and incoherent vector meson production processes from heavy nuclei~\cite{Accardi:2012qut}. This measurement provides a means of probing the gluon distribution within an ion and is particularly sensitive to saturation phenomena (see, for example,~\cite{Toll:2012mb}). While both coherent and incoherent processes are of great interest, measuring the coherent production processes at high momentum transfer ($t$) or measuring the incoherent production process at a very small momentum transfer presents significant challenges. The discrimination of incoherent processes is achievable due to the extensive kinematic coverage provided by the far-forward (FF) detector array. 

In \cite{Chang:2021jnu}, the authors explored the vetoing capabilities of  FF detectors to distinguish between coherent and incoherent production processes. Incoherent processes are characterized by ion breakup and the production of ion fragments, including free protons and neutrons, in the forward direction. Protons were effectively vetoed by the B0 tracker, OMD, and RP detectors, while the B0 EMCAL and ZDC vetoed neutral particles (such as photons and neutrons). As a result, most events that passed the full event selection criteria were identified as incoherent events, primarily through ion de-excitation. The authors demonstrated that the vetoing techniques suppressed incoherent processes by two orders of magnitude, leading to a signal-to-background ratio above unity for events with $t$ values near the first diffractive minimum. 

Additionally, most of the remaining background events originate from incoherent processes involving ion de-excitation, where the emitted photons escape detection. These processes, which have never been systematically studied, offer a promising new area of investigation with potential implications for the EIC physics program.

\subsubsection*{Virtual Compton Scattering (u-channel)}
While the DVCS is also considered one of the ``golden channels'' of the EIC physics program due to its clear interpretation in terms of generalized Parton distributions~\cite{Burkardt:2002hr},  virtual Compton backward scattering (u-channel) involves a large momentum transfer and may play a significant role in baryon stopping in heavy-ion collisions. The main background for this process is a coherent $\pi^0$ production, where $\pi^0\to\gamma\gamma$ can be misinterpreted as a single photon due to a small scattering angle or if one of the photons escapes detection~\cite{Sweger:2023bmx}. 

The highly segmented ZDC provides powerful discrimination across all energy ranges. The angular separation of the two photons at the beam energy $E_\textrm{beam}$ at the distance of the ZDC from the IP is given by $\Delta x^{\gamma\gamma} = 70\cdot m_\pi /E_\textrm{beam}$ meters, which corresponds to separations of 23~cm, 9.5~cm, and 3.4~cm for beam energies of 41, 100, and 275 GeV, respectively. The authors of~\cite{Sweger:2023bmx} demonstrated that the $\pi^0$ background is reduced to a few percent in $ep$ collisions at $18\times 275~\textrm{GeV}^2$. Moreover, the inclusion of the B0 detector will enhance signal acceptance by a factor of two or ten at a beam energy of 100 and 41 GeV.

\section{Summary}
Comprehensive acceptance studies for all far-forward and far-backward detectors have been conducted, and their performance is well understood based on currently available information. The extended far-forward detector array allows an impressive extension of the nominal physics program~\cite{AbdulKhalek:2021gbh} foreseen with the existing detectors. The focus has now shifted to simulation studies of various processes in preparation for the ePIC Technical Design Report.

\end{document}